\documentclass[a4paper,fleqn]{cas-dc}

\usepackage[numbers]{natbib}

\usepackage[capitalize]{cleveref}
\usepackage{upgreek}
\usepackage{float}
\usepackage{algorithm}
\usepackage{algorithmic}
\newcommand{\SubState}{\STATE \hspace{\algorithmicindent}}
\usepackage{hyperref}
\hypersetup{
    colorlinks=true,        
    linkcolor=blue,        
    citecolor=blue,        
    urlcolor=blue,         
    filecolor=blue        
}
\usepackage{enumitem} 
\begin{document}
\let\WriteBookmarks\relax
\def\floatpagepagefraction{1}
\def\textpagefraction{.001}
\shorttitle{Joint-decoupled iterative CBCT reconstruction with hybrid scatter estimation and voxel-adaptive beam hardening correction}
\shortauthors{J.~Sun et al.}

\title[mode=title]{Joint-decoupled iterative CBCT reconstruction with hybrid scatter estimation and voxel-adaptive beam hardening correction}
\tnotemark[1,2]

\tnotetext[1]{This work was supported by the National Natural Science Foundation of China, Grant/Award Number 12426308, by the National Key Research and Development Program of China, Grant/Award Numbers 2020YFA0712200 and 2023YFA1011402, and by the Postdoctoral Research Fund of Inner Mongolia University, Grant/Award Number 10000-A25206032.}

\tnotetext[2]{J. Sun and J. M. L\'{e}tang contributed equally to this work.}

\author[1,2]{Jianing Sun}[orcid=0009-0008-4218-0923]
\ead{jn.sun@cnu.edu.cn}
\credit{Methodology, Software, Writing – original draft, Writing – review \& editing, Formal analysis, Data curation}

\affiliation[1]{organization={College of Mathematics Science, Inner Mongolia Normal University},
    city={Hohhot},
    postcode={010022},
    country={China}}
	
\affiliation[2]{organization={School of Mathematical Sciences, Capital Normal University},
    city={Beijing},
    postcode={100048},
    country={China}}

\affiliation[3]{organization={CREATIS (CNRS UMR 5220, Inserm U1294), INSA-Lyon, Universit\'{e} Claude Bernard Lyon 1},
    city={Lyon},
    postcode={69373},
    country={France}}

\affiliation[4]{organization={School of Data Science and Information Technology, China Women's University},
    city={Beijing},
    postcode={100101},
    country={China}}
    
\affiliation[5]{organization={  College of Computer Science (College of Software), Inner Mongolia University},
    city={Hohhot},
    postcode={010021},
    country={China}}
    
\author[3]{Jean Michel L\'{e}tang}[orcid=0000-0003-2583-782X]
\ead{jean.letang@creatis.insa-lyon.fr}
\credit{Methodology, Software, Writing – original draft, Writing – review \& editing, Formal analysis, Investigation}

\author[4]{Qixiang Sun}[orcid=0000-0001-7288-4501]
\ead{2501007@cwu.edu.cn}
\credit{Software, Data curation, Writing – review \& editing}

\author[2]{Guangyin Li}[orcid=0009-0003-1085-4798]
\ead{kwongyamlie1999@gmail.com}
\credit{Software, Data curation, Writing – review \& editing}

\author[5]{Ligen Shi}[orcid=0009-0005-5990-0606]
\ead{ligenshi0826@gmail.com}
\credit{Resources, Writing – review \& editing, Formal analysis}

\author[2]{Xing Zhao}[orcid=0000-0003-2394-9852]
\cormark[1]
\ead{zhaoxing_1999@126.com}
\credit{Methodology, Supervision, Writing - review \& editing, Resources, Funding acquisition}
\cortext[1]{Corresponding author}

\begin{abstract}
\noindent\textbf{Background and Objective:} Cone-beam computed tomography (CBCT) is fundamentally challenged by scatter and beam hardening artifacts, which originate from X-ray scattering and the polychromatic nature of the X-ray spectrum, respectively. These two types of artifacts are intricately coupled in reconstructed images and manifest with similar streaking and cupping features, severely compromising high-precision CBCT imaging.

\noindent\textbf{Methods:} This paper proposes a physics-driven iterative framework rooted in the polychromatic Polyquant attenuation model, which decouples these artifacts by establishing an optimization loop between scatter estimation and relative electron density (RED) reconstruction. We develop a hybrid strategy for scatter estimation, in which the first-order scattering component is analytically derived based on a polychromatic physical model to preserve high-frequency structural information, whereas the smoother multiple scattering component is efficiently estimated via an object-adaptive convolution module. Subsequently, for beam-hardening correction, we introduce a voxel-adaptive update mechanism that solves linearized, scatter-corrected polychromatic equations to derive optimal weights, enabling direct RED refinement without manual parameter tuning.

\noindent\textbf{Results:} The proposed method was validated through comprehensive studies on biomedical phantoms, utilizing both Monte Carlo simulations and physical experiments. Representative results demonstrate that the proposed method outperforms state-of-the-art techniques, with the mean relative error decreased from 11.96\% to 1.27\% for the anthropomorphic head phantom and from 12.55\% to 5.46\% for the physical Yin-Yang phantom.

\noindent\textbf{Conclusions:} This study presents a physics-driven iterative framework that effectively disentangles scatter and beam-hardening artifacts by synergizing a hybrid scatter estimation strategy with a voxel-adaptive update mechanism for beam-hardening-free RED reconstruction, enabling high-fidelity diagnostic imaging.

\end{abstract}




\maketitle

\section{Introduction}
Cone-beam computed tomography (CBCT) has emerged as a vital technology for image-guided radiation therapy \cite{posiewnik2019review} and dental imaging \cite{kaasalainen2021dental}, offering a larger field of view, higher spatial resolution, and lower cost \cite{deng2024multi}. However, its clinical utility is compromised by artifacts stemming from two inherent physical phenomena \cite{huang2026coupled}. First, the large cone-beam irradiation geometry creates a substantial scattering volume, resulting in excessive scatter signals reaching the detector \cite{mason2018quantitative}. For instance, experiments on our in-house CBCT system reveal that the scatter-to-primary ratio (SPR) can reach approximately 3.0 for a Yin-Yang phantom (15 cm diameter, 6 cm height) composed of water- and bone-equivalent inserts. Second, beam hardening arises from the preferential attenuation of lower-energy photons within the polychromatic X-ray spectrum as they traverse the object. Consequently, conventional reconstruction algorithms relying on monoenergetic assumptions, such as filtered back-projection (FBP) or the simultaneous algebraic reconstruction technique (SART), yield images corrupted by these combined artifacts \cite{grimmer2011empirical,wu2015iterative}. Compounding this challenge, scatter and beam-hardening artifacts are intricately coupled and manifest with visually similar cupping and streaking patterns \cite{lee2015single}, making them virtually indistinguishable in the reconstructed images. This degradation not only impairs diagnostic image quality but also hinders the quantitative accuracy essential for critical applications, such as dose calculation in radiation therapy \cite{zhu2009scatter} and myocardial perfusion assessment \cite{kitagawa2010characterization}.

Given that scatter signals severely distort the measured projection intensities, the fidelity of these measurements is an important prerequisite for effective beam-hardening correction (BHC). Without effective scatter removal, the underlying physical assumptions required for hardening compensation remain fundamentally compromised. Therefore, accurate scatter estimation is widely regarded as the cornerstone of high-fidelity CBCT reconstruction. Existing scatter correction strategies are generally categorized into hardware-based and software-based approaches \cite{Ruehrnschopf2011, ruhrnschopf2011general}. Hardware-based methods, such as the air-gap technique \cite{groedel1926bedeutung, persliden1997scatter}, anti-scatter grids \cite{neitzel1992grids}, beam stop arrays (BSA) \cite{ning2004x}, and primary modulators \cite{zhu2006scatter, gao2010scatter}, aim to physically reject or measure scattered radiation. However, their clinical utility is often limited by inherent trade-offs, such as geometric constraints (air-gaps), degradation of the signal-to-noise ratio and secondary artifacts (grids), increased radiation dose (beam stops), or strict manufacturing requirements (modulators). Driven by the constraints of hardware solutions, software-based scatter correction has become a focal point of research. These strategies are generally categorized into Monte Carlo (MC) simulations \cite{badal2009accelerating, zhang2020scatter,xu2025accelerated}, deep learning (DL) techniques \cite{maier2018deep, roser2021x, zhuo2023scatter}, and model-based approaches \cite{love1987scatter, freud2004deterministic,sun2010improved,bhatia2016scattering, gong2017physics, mason2018quantitative}. Monte Carlo (MC) simulations are widely recognized as the gold standard for scatter estimation, yet they entail a heavy computational burden arising from the need to simulate massive numbers of particle interactions \cite{pointon2023simulation}. Although acceleration strategies such as variance reduction and GPU parallelization have been developed, their efficiency in clinical scenarios relies on high-performance computing infrastructure. Deep learning (DL) techniques have emerged as a promising alternative, offering rapid inference speeds. However, these data-driven models often lack physical interpretability and can lead to unpredictable artifacts. Furthermore, their clinical applicability is fundamentally constrained by the scarcity of high-quality paired training data and limited generalization across different scanners and imaging protocols. Model-based approaches are generally categorized into convolution-based and analytical strategies. While convolution-based methods offer high computational efficiency, they typically rely on kernel approximations. Consequently, even adaptive variants like FASKS \cite{sun2010improved} struggle to accurately capture the high-frequency variations of first-order scattering at highly heterogeneous interfaces, particularly the sharp bone-tissue transitions of the skull. Alternatively, analytical strategies model scatter based on physical principles, theoretically enabling the accurate characterization of first-order scattering. Building on this premise, the physics-based multi-source CT scatter correction (PMSC) \cite{gong2017physics} adapts this strategy to CBCT, where the exact material composition is unknown. It employs an iterative framework that uses scatter-corrected projections to reconstruct the object’s linear attenuation coefficient (LAC) distribution at a mean effective energy, and subsequently utilizes this updated map to refine the scatter estimation. However, in practice, PMSC compromises physical accuracy through three simplifications, including: 1) neglecting the contribution of multiple scattering by approximating the total scatter signal solely as first-order scattering, 2) overlooking the energy dependence of first-order scatter intensity by relying on the reconstructed LAC image at the mean effective energy, and 3)  incorporating bias into the estimated first-order scattering by utilizing the LAC image compromised by beam-hardening artifacts. 

Beyond the limitations of individual scatter estimation methods, isolated correction is fundamentally insufficient due to the inherent coupling between scatter and beam-hardening artifacts. To resolve this dependency, existing strategies are generally categorized into image-domain post-processing \cite{lou2013joint, wu2015iterative}, multi-spectral decomposition \cite{lewis2022energy, deng2024multi, lorenzon2024joint}, and projection-domain joint correction \cite{grimmer2011empirical, mason2018quantitative}. Image-domain post-processing suppresses artifacts by enforcing tissue uniformity, typically retaining high-frequency streaks. Multi-spectral decomposition simultaneously corrects scatter and reconstructs basis materials, yet necessitates multi-energy data acquired via photon-counting detectors or multiple scans at distinct voltages. Projection-domain joint correction estimates scatter intensity from measured projections by accounting for the polychromatic CBCT forward projection, and utilizes the corrected data to reconstruct images approximately free of scatter and beam-hardening artifacts. 
A representative method, Polyquant-PolySKS \cite{mason2018quantitative}, derives a polychromatic scatter kernel superposition (PolySKS) algorithm based on the Polyquant attenuation model \cite{Mason_2017} and incorporates it into an alternating scheme to iteratively reconstruct relative electron density (RED) images via the Prox-Polyquant BHC algorithm. 
However, it still faces several challenges in removing scatter and beam-hardening artifacts, including: 1) scatter estimation inaccuracy at highly heterogeneous interfaces due to the limited capability of convolution kernel approximations in modeling complex high-frequency scatter signals, which subsequently degrades the RED image reconstruction; and 2) slow convergence and cumbersome parameter tuning in the BHC, attributed to the lack of precise step-size calculation and the requirement for manual adjustment of regularization parameters.

\begin{figure}[pos=b] 
\centering
\includegraphics[width=\linewidth]{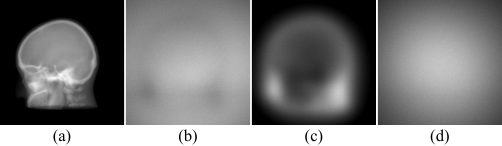}
\caption{Visual comparison of different scatter components of a CBCT skull radiography. (a) Lateral view in the negative logarithm domain (display window is [0, 6.8]); (b) normalized Compton scattering; (c) normalized Rayleigh scattering; and (d) normalized multiple scattering. The display window for (b)-(d) is [0.2, 1.2].}
\label{Introduction-Analysis}
\end{figure}

To address these challenges, this paper proposes a physics-driven iterative reconstruction framework built upon the Polyquant attenuation model, serving as the foundation for 
scatter estimation and 
parameter-free BHC. For effective scatter estimation, our proposed strategy is motivated by the notable observation in \cref{Introduction-Analysis} that first-order scattering (including Compton and Rayleigh components) preserves high-frequency structural information, whereas multiple scattering manifests as a diffuse, slowly varying distribution. 
Leveraging these distinct characteristics, we introduce a hybrid scatter estimation strategy that significantly improves scatter correction in structurally complex regions compared to standard convolution models, such as PolySKS, while effectively addressing the limitations of the PMSC algorithm. Specifically, for first-order scattering, we derive a polychromatic analytical solution that explicitly calculates the interaction probabilities for each voxel using the iteratively updated RED image and the corresponding Compton and Rayleigh cross-sections. Subsequently, a ray-tracing technique accumulates these single-scatter contributions at the detector. For multiple scattering, we employ an object-adaptive convolution module for rapid estimation, where the scatter distribution is modeled by convolving the measured projections with an object-specific kernel.
In the subsequent BHC step, we propose a voxel-adaptive mechanism to reconstruct the RED image, effectively addressing the update limitations of the Prox-Polyquant method. To achieve this, we linearize the nonlinear polychromatic forward model at the current RED estimate. Then, SART is employed to solve the linearized, scatter-corrected polychromatic projection equations, thereby deriving optimal update weights. Overall, the proposed physics-driven iterative framework ensures accurate scatter estimation and enables the rapid reconstruction of CBCT images that are visually free from scatter and beam-hardening artifacts. The main contributions of this paper are as follows: 
\begin{enumerate}[label=\arabic*)]
    \item We propose a physics-driven iterative reconstruction framework, effectively decoupling and eliminating scatter and beam-hardening artifacts based on fundamental physical principles.
    \item We propose a hybrid scatter estimation strategy leveraging distinct frequency characteristics that addresses the inaccuracy of convolution models in representing high-frequency scatter signals, while overcoming the limitations of existing PMSC algorithms, such as the monochromatic assumption and the neglect of multiple scattering.
    \item We develop a voxel-adaptive BHC mechanism that overcomes the limitations of the Prox-Polyquant method regarding update step sizes and manual parameter tuning, enabling robust and rapid image reconstruction.
\end{enumerate}

\section{Methods}
The Polyquant model \cite{Mason_2017} establishes a data-driven mapping between RED $\boldsymbol{x}$ and LACs through a piecewise linear function. For a given energy $E$, the LAC map $\boldsymbol{\mu}(\boldsymbol{x},E)$ is formulated  as: 
 \begin{equation}\label{Polyquant}
	\boldsymbol{\mu}(\boldsymbol{x},E) = \sum_{i=1}^{N_{f}} \boldsymbol{f}_{i}(\boldsymbol{x})\odot\left(\alpha_{i}(E)\,\boldsymbol{x} + \beta_{i}(E)\right),
\end{equation}
where $\boldsymbol{f}_{i}$ denotes the binary class indicator for material category $i$, $\alpha_{i}(E)$ and $\beta_{i}(E)$ are the energy-dependent model parameters, and $\odot$ represents the Hadamard (element-wise) product. The two-component model ($N_{\!f} = 2$) is well-suited for reconstructing water-bone compositions, with water and bone serving as the basis materials. The framework can be extended to scenarios involving metallic implants or contrast agents by setting $N_{\!f} = 3$.

In the single-source CBCT imaging system, the measured total projection intensity $I^t$ at position $\boldsymbol{u}$ consists of primary transmission $I^{p}$ and forward scatter $I^s$. The relationship is expressed as:
\begin{equation}
    I^t(k,\boldsymbol{s},\boldsymbol{u}) = I^p(k,\boldsymbol{s},\boldsymbol{u}) + I^s(k,\boldsymbol{s},\boldsymbol{u}),
\end{equation}
where 
the integral path is from the X-ray source $\boldsymbol{s}$ to a detector pixel $\boldsymbol{u}$ at the $k^{\text{th}}$ projection angle. For brevity, we will omit the subscript $k$ in subsequent equations when the context is clear.
Considering a polychromatic X-ray imaging process based on the Polyquant attenuation model, 
the primary transmission $ I^p$ is described as:
\begin{equation}\label{polyquant}
I^{p}\left(\boldsymbol{s}, \boldsymbol{u}\right) = \sum_{E}I^p\left(\boldsymbol{s}, \boldsymbol{u}, E\right)
\end{equation}
with (using $^{\top}$ for the transpose operator)
\begin{equation}\label{eq:Ip-su}
\hspace*{-0.5cm}
\mbox{\small$\displaystyle
I^{p}\left(\boldsymbol{s}, \boldsymbol{u}, E\right) = I^{0}(\boldsymbol{s}, \boldsymbol{u})\, q(E)\,D(E) \exp\left(-\boldsymbol{R}^{\top}_{\boldsymbol{s}, \boldsymbol{u}}\boldsymbol{\mu}_{\boldsymbol{s},\boldsymbol{u}}(\boldsymbol{x},E)\right),
$}
\end{equation}
where $I^{0}(\boldsymbol{s}, \boldsymbol{u})$ denotes the photon intensity emitted from the X-ray source $\boldsymbol{s}$ and reaching detector pixel $\boldsymbol{u}$ when no object is present, $q(E)$ is the photon fraction of the incident X-ray spectrum in energy bin $E$ (the integral of $q$ over the energy bins is unity and we assume the emitted spectrum is spatially invariant), and $D(E)$ represents the energy-dependent response function of the detector. The combined term $q(E)D(E)$ represents the normalized detector-weighted spectrum for bin $E$.
The vector $\boldsymbol{R}_{\boldsymbol{s}, \boldsymbol{u}}$ 
(resp.~$\boldsymbol{\mu}_{\boldsymbol{s},\boldsymbol{u}}(\boldsymbol{x},E)$)
contains the intersection lengths (resp.~the LAC) in each of the voxels from the source  $\boldsymbol{s}$ to the detector pixel $\boldsymbol{u}$.

Our objective is to accurately reconstruct the RED image $\boldsymbol{x}$ from the measured projection intensities $I^t$, which are composed of both a primary component and a scatter signal. The primary signal is related to the unknown RED image via a non-linear polychromatic projection equation, as described in \cref{polyquant}, while the scatter distribution depends on its scattering cross-section. To address this coupled problem, we propose an iterative framework that alternately estimates the scatter signal and updates the RED image using scatter-corrected polychromatic projections.

\subsection{Analytic module for first-order scattering}

\begin{figure}[pos=t]
\centering
\includegraphics[width=\linewidth]{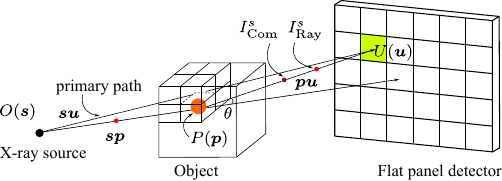}
\caption{The schematic diagram of the proposed polychromatic first-order scattering module.}
\label{figure_first-order scattering}
\end{figure}

\cref{figure_first-order scattering} illustrates a schematic of the proposed first-order scattering module. Photons originating from the X-ray source point $O$ (at position $\boldsymbol{s}$) travel along a primary path segment ${OP}$.
At a scattering voxel $P$ (at position $\boldsymbol{p}$), they are deflected by an angle $\theta$ and propagate along a secondary path segment ${PU}$, experiencing further attenuation before reaching the detector point $U$ (at position $\boldsymbol{u}$). 
The modeling of this process begins with the primary beam. For a mono-energetic beam, the primary transmission with energy $E$ reaching voxel $P$ can be expressed as \cref{eq:Ip-su} with voxel $\boldsymbol{p}$ replacing detector pixel $\boldsymbol{u}$:
\begin{equation}    
    I^{p}(\boldsymbol{s},\boldsymbol{p},E) =  I^{0}(\boldsymbol{s}, \boldsymbol{p})\,q(E)\,\exp\left(-\boldsymbol{R}^{ \top}_{\boldsymbol{s}, \boldsymbol{p}}\boldsymbol{\mu}_{\boldsymbol{s},\boldsymbol{p}}(\boldsymbol{x},E)\right),
\end{equation}
where the detector response $D(E)$ no longer intervenes in the equation.
Subsequently, we consider the scattering intensity $I^s(\boldsymbol{p},\boldsymbol{u},E)$ with energy $E$ from voxel $\boldsymbol{p}$ to detector pixel $\boldsymbol{u}$, which is proportional to the incident primary transmission from source $\boldsymbol{s}$ and the differential probability of scattering.
This total differential probability is the sum of the contributions from Compton scattering $\mathrm{d}p^{\text{Com}}(\boldsymbol{p}, E, \theta)$ and Rayleigh scattering $\mathrm{d}p^{\text{Ray}}(\boldsymbol{p}, E, \theta)$, but these differential cross sections require knowledge of the atomic number and molar mass of the material, and only the RED $\boldsymbol{x}$, i.e., the reconstructed value, is available. We propose making an approximation using values relative to water when necessary.
This gives:
\begin{subequations} 
\begin{small} %
\begin{align}
\hspace*{-0.3cm} 
\mathrm{d}p^{\text{Com}}(\boldsymbol{p}, E, \theta) &= \frac{\uprho_{\boldsymbol{p}}(\boldsymbol{x}){{N}_{\!A}}\mathrm{d}V}{M_{\text{w}}{\mathrm{d}A}}\frac{\mathrm{d}\sigma_{\text{C}}(E, \theta)}{\mathrm{d}\Omega}S_{\text{w}}(E,\theta)\mathrm{d}\Omega_{\boldsymbol{p}},
\\
\hspace*{-0.3cm} 
\mathrm{d}p^{\text{Ray}}(\boldsymbol{p}, E, \theta) &= \frac{\uprho_{\boldsymbol{p}}(\boldsymbol{x}){{N}_{\!A}}\mathrm{d}V}{M_{\text{w}}{\mathrm{d}A}}\frac{\mathrm{d}\sigma_{\text{T}}(E, \theta)}{\mathrm{d}\Omega}F_{\text{w}}^2(E,\theta)\mathrm{d}\Omega_{\boldsymbol{p}},
\end{align}
\end{small}
\end{subequations}%
\par\noindent The terms in these expressions are defined as follows: $\uprho_{\boldsymbol{p}}$ represents the mass density at voxel $\boldsymbol{p}$, which is determined via linear interpolation from a pre-established lookup table mapping relative electron density $\boldsymbol{x}$ to the mass densities of reference materials; $N_A$ is the Avogadro constant; $M_{\text{w}}$ is the molecular molar mass of water; $\mathrm{d}V$ is voxel volume; $\mathrm{d}A$ is the cross-section area of the incident beam; and $\mathrm{d}\Omega_{\boldsymbol{p}}$ stands for the solid angle subtended by the detector element $\boldsymbol{u}$ from the voxel $\boldsymbol{p}$. The terms $S_{\text{w}}(E,\theta)$ and $F_{\text{w}}^{2}(E,\theta)$ are the molecular incoherent scattering function and coherent scattering form factor of water, respectively. Both are typically calculated via a weighted summation of the corresponding atomic scattering factors \cite{tartari1999molecular, hubbell1975atomic}.
The angular distribution of scattered photons is described by their respective differential cross-sections. For Compton (incoherent) scattering, the Klein-Nishina formula \cite{hubbell1975atomic} describes the differential cross-section, and the scattered photon energy $E_{\text{Com}}$ is reduced as follows:
\begin{small}
\begin{subequations} 
\begin{align}
    \begin{split}
    \frac{\mathrm{d}\sigma_{\text{C}}\left(E,\theta \right)}{\mathrm{d}\Omega} &= \frac{r_e^2}{2} \left[1 + \varepsilon(1 - \cos\theta)\right]^{-2} \\
    & \quad \left( 1 + \cos^2\theta + \frac{\varepsilon^2(1-\cos\theta)^2}{1 + \varepsilon(1-\cos\theta)} \right),
    \end{split}    
    \\
    E_{\text{Com}} &= \frac{E}{ 1 + \varepsilon( 1 - \cos\theta)},
\end{align}
\end{subequations}
\end{small}
\par \noindent  where $r_e$ is the classical electron radius, $ \varepsilon = E/{511 \text{keV}}$ represents the incident photon energy in units of electron rest mass. For Rayleigh (coherent) scattering, the process is described by the Thomson differential cross-section \cite{hubbell1975atomic} and the photon energy $E_{\text{Ray}}$ is conserved:
\begin{subequations} 
\begin{align}
    \frac{\mathrm{d}\sigma_{\text{T}}\left(E,\theta \right)}{\mathrm{d}\Omega} &= \frac{r_e^2}{2} \,\left(1 + \cos^2(\theta)\right),\\
    E_{\text{Ray}} &= E.
\end{align}
\end{subequations} 
The photons scattered at $\boldsymbol{p}$ are further attenuated. The intensity arriving at the detector pixel $\boldsymbol{u}$, for the incident energy $E$, is the sum of Compton and Rayleigh components:
\begin{small}
\begin{align}
    \hspace*{-0.8cm}
    I_{\text{fir}}^s(\boldsymbol{p},\boldsymbol{u}, E) = I_{\text{Com}}^{s}(\boldsymbol{p},\boldsymbol{u},E)\,D(E_{\text{Com}}) + I_{\text{Ray}}^{s}(\boldsymbol{p},\boldsymbol{u},E)\,D(E_{\text{Ray}}),
\end{align}
\end{small}
with 
\begin{subequations}
\begin{align}
    \begin{split}
    I_{\text{Com}}^{s}(\boldsymbol{p},\boldsymbol{u},E) = \,& I^{p}(\boldsymbol{s},\boldsymbol{p},E)\,\mathrm{d}p^{\text{Com}}(\boldsymbol{p},E,\theta)\\&\mathrm{exp}\left(-\boldsymbol{R}^{\top}_{\boldsymbol{p},\boldsymbol{u}}\boldsymbol{\mu}_{\boldsymbol{p},\boldsymbol{u}}\left(\boldsymbol{x},E_{\text{Com}}\right) \right),
    \end{split}\\
    \begin{split}
    I_{\text{Ray}}^{s}(\boldsymbol{p},\boldsymbol{u},E) = \,& I^{p}(\boldsymbol{s},\boldsymbol{p},E)\,\mathrm{d}p^{\text{Ray}}(\boldsymbol{p},E,\theta)\\&\mathrm{exp}\left(-\boldsymbol{R}^{\top}_{\boldsymbol{p},\boldsymbol{u}}\boldsymbol{\mu}_{\boldsymbol{p},\boldsymbol{u}}\left(\boldsymbol{x},E_{\text{Ray}}\right) \right).
    \end{split}
\end{align}
\end{subequations}
Finally, the total single scattering at the detector pixel $\boldsymbol{u}$ is obtained by integrating the contributions from all scattering voxels $\boldsymbol{p}$ within the object and over the entire energy spectrum:
\begin{small}
\begin{align}\label{firOrder_scattering}
    \hspace*{-0.83cm}
    I_{\text{fir}}^s(\boldsymbol{s},\boldsymbol{u}) = \sum_{E}\sum_{\boldsymbol{p}} \left( I_{\text{Com}}^{s}(\boldsymbol{p},\boldsymbol{u},E) D(E_{\text{Com}}) + I_{\text{Ray}}^{s}(\boldsymbol{p},\boldsymbol{u},E) D(E_{\text{Ray}}) \right).
\end{align}
\end{small}

\subsection{Object-adaptive convolution module for multiple scattering}
Compared with first-order scattering, multiple scattering events constitute a smoother, low-frequency component. This physical property makes the multiple scattering $I_{\text{mul}}^s$ amenable to a convolution model. Accordingly, we approximate it as  the convolution of the measured projection intensity $I^{t}$ with a scatter kernel $h^{s}$:
\begin{equation} \label{multiple scattering} 
I^{s}_{\text{mul}}(\boldsymbol{s},\boldsymbol{u}) = I^{t}(\boldsymbol{s},\boldsymbol{u}) \ast h^{s}(\boldsymbol{u}),
\end{equation} 
where $\ast$ denotes the convolution operator. The scatter kernel $h^{s}$ is composed of an amplitude factor $c$ and a double Gaussian kernel $g$ \cite{ohnesorge1999efficient}:
\begin{equation}
    h^s(\boldsymbol{u}) = c(\boldsymbol{u})\,g(\boldsymbol{u}).
\end{equation}
The forward amplitude factor $c$ is defined as:
\begin{equation}
    c\left(\boldsymbol{u}\right) = A\,\left(\frac{I^{t}\left(\boldsymbol{s},\boldsymbol{u}\right)}{I^{0}\left(\boldsymbol{s},\boldsymbol{u}\right)}\right)^{\alpha} \left(\log\left( \frac{I^{0}\left(\boldsymbol{s},\boldsymbol{u}\right)}{I^{t}\left(\boldsymbol{s},\boldsymbol{u}\right)}   \right)  \right)^{\beta},
\end{equation}
where $A$, $\alpha$ and $\beta$ are parameters that influence the amplitude of the scatter kernel $h^{s}$. The kernel $g$ consists of two symmetric Gaussian kernels \cite{suri2006comparison}, with standard deviation $\sigma_1, \sigma_2$ and weighting factor $B$:
\begin{align}
    g(\boldsymbol{u}) &= \exp\left(- \frac{\boldsymbol{u}^{\top}\boldsymbol{u}}{2\sigma_1^2}\right)
    + B \exp\left(- \frac{\boldsymbol{u}^{\top}\boldsymbol{u}}{2\sigma_2^2} \right).
\end{align}

The complete set of parameters is defined by the vector $\boldsymbol{\phi} = \left(A, B, \alpha, \beta, {\sigma}_1, {\sigma}_2\right)$, adaptively calibrated by leveraging prior scatter information specific to the scanned object.
The reconstructed image $\boldsymbol{x}$ is first converted into a mass density map to generate reference distributions. This map then serves as the input to a GPU-based open-source MC package (MC-GPU) \cite{badal2009accelerating}, allowing for the accurate modeling of X-ray photon interactions within the scanned object. To ensure efficiency without sacrificing accuracy, simulations are performed only at a sparse set of angles, using the high similarity of multiple scattering distributions between adjacent views. MC-GPU outputs the total projection intensity $I^{t}_{\text{MC}}$ and multiple scattering component $I^{s}_{\text{MC},\text{mul}}$, which are used to calibrate parameter vector $\boldsymbol{\phi}$. This calibration is achieved by minimizing the least-squares error between the module's estimate and the simulated ground truth.

\subsection{Voxel-adaptive module for beam hardening}
In practice, scatter intensity can significantly exceed the primary signal. During an iterative process, subtracting the scatter estimate from the measured projection intensity can yield a non-physical negative primary signal estimate. To circumvent this, we employ a multiplicative operation instead of subtraction \cite{bhatia2016scattering}.
The primary signal estimate is initialized with the measured total projection:
\begin{equation}    
I^{p,0}_{\text{sc}}\left(\boldsymbol{s},\boldsymbol{u}\right) = I^{t}\left(\boldsymbol{s},\boldsymbol{u}\right).
\end{equation}
To obtain the estimate for the $(n+1)$-th iteration, given the  reconstructed relative electron density (RED) image $\boldsymbol{x}^{n}$, we estimate the primary transmission with scatter correction $I^{p,n+1}_{\text{sc}}$ as follows:
\begin{equation}
    I^{p, n+1}_{\text{sc}}(\boldsymbol{s},\boldsymbol{u}) = \frac{I^{p,n}_{\text{sc}}(\boldsymbol{s},\boldsymbol{u}) \odot 
    I^{t}(\boldsymbol{s},\boldsymbol{u})}{I^{p,n}_{\text{sc}}(\boldsymbol{s},\boldsymbol{u}) + I_{\text{fir}}^{s,n}(\boldsymbol{s},\boldsymbol{u}) + I^{s}_{\text{mul}}(\boldsymbol{s},\boldsymbol{u})}.
    \label{corrPrimary}
\end{equation}
The terms $I_{\text{fir}}^{s,n}$ and $I^{s,n}_{\text{mul}}$ denote the estimated first-order and multiple scattering components, respectively. These are derived from the reconstructed RED image $\boldsymbol{x}^{n}$ as detailed in \cref{firOrder_scattering,multiple scattering}.
Subsequently, to mitigate beam-hardening artifacts, the scatter-corrected intensity $I^{p, n+1}_{\text{sc}}$ is integrated into the forward model. This formulation is established in the negative logarithmic domain to leverage the approximate linearity of the polychromatic projections:
\begin{small}
\begin{equation}
    P^{p, n+1}_{\text{sc}}\left(\boldsymbol{s},\boldsymbol{u}\right) = -\log\left(\frac{I^{p, n+1}_{\text{sc}}(\boldsymbol{s},\boldsymbol{u})}{I^{0}(\boldsymbol{s},\boldsymbol{u})}\right), \label{corrProj}
\end{equation}
\end{small}
\par \noindent where $P^{p, n+1}_{\text{sc}}$ is the estimated scatter-corrected projection at the $(n+1)$-th iteration. From \cref{eq:Ip-su} we also have
\begin{small}
\begin{equation}    
    P^{p, n+1}\left(\boldsymbol{s},\boldsymbol{u}\right) = -\log\left(\sum_{E} q(E)\,D(E)\,\exp\left(-\boldsymbol{R}^{\top}_{\boldsymbol{s},\boldsymbol{u}}\boldsymbol{\mu}_{\boldsymbol{s},\boldsymbol{u}}(\boldsymbol{x}^{n+1},E)\right)\right).\label{corrPolyquant} 
\end{equation}
\end{small}
\par \noindent To compute the image update, we linearize the nonlinear relationship in \cref{corrPolyquant} via a first-order Taylor expansion around the current estimate $\boldsymbol{x}^{n}$ (cf. \cref{Polyquant}):
\begin{equation}
    P^{p, n+1}\left(\boldsymbol{s},\boldsymbol{u}\right) = P^{p, n}\left(\boldsymbol{s},\boldsymbol{u}\right) + \frac{\boldsymbol{T}^{n}\left(\boldsymbol{s},\boldsymbol{u}\right)}{b^{n}\left(\boldsymbol{s},\boldsymbol{u}\right)} \Delta \boldsymbol{x}^{n} + o(\lVert \Delta \boldsymbol{x}^{n}  \rVert),
    \label{Taylor}
\end{equation}
with 
\begin{small}
\begin{subequations}
\begin{align}\begin{split}
    \boldsymbol{T}^{n}\left(\boldsymbol{s},\boldsymbol{u}\right) &= \sum_{E} q(E)D(E) \\
    & \sum_{i=1}^{N_f}\boldsymbol{R}_{\boldsymbol{s},\boldsymbol{u}}^{\top} \boldsymbol{f}_{i}(\boldsymbol{x}^n)\,\alpha_{i}(E)\,\exp\left(-\boldsymbol{R}^{\top}_{\boldsymbol{s},\boldsymbol{u}}\boldsymbol{\mu}_{\boldsymbol{s},\boldsymbol{u}}(\boldsymbol{x}^n,E)\right),
    \end{split}
    \\ 
    b^{n}\left(\boldsymbol{s},\boldsymbol{u}\right) &= \sum_{E} q(E)\,D(E)\exp \left(-\boldsymbol{R}^{\top}_{\boldsymbol{s},\boldsymbol{u}}\boldsymbol{\mu}_{\boldsymbol{s},\boldsymbol{u}}(\boldsymbol{x}^n,E)\right).
\end{align}
\end{subequations}
\end{small}

Iterative algebraic methods work well for solving the linear equation with multiple unknowns as in \cref{Taylor}. Among these, we adopt the simultaneous algebraic reconstruction technique (SART) due to its superior convergence rate and robustness to noise \cite{andersen1984simultaneous}. SART is an iterative angle-by-angle updating strategy, using corrections computed simultaneously from a subset of projections. The update rule for the $j$-th voxel at the $(n+1)$-th iteration is given by:
\begin{small}
\begin{subequations}
 \begin{align}\hspace*{-1em} x^{n+1}_{j} &= x^{n}_{j} + \sum_{\boldsymbol{u}} w_{\boldsymbol{s},\boldsymbol{u},j}\frac{b^{n}\left( P^{p, n+1}_{\text{sc}}(\boldsymbol{s},\boldsymbol{u}) - P^{p, n}(\boldsymbol{s},\boldsymbol{u}) \right)T_{j}^{n}}{\lVert \boldsymbol{T}^{n} \rVert^2}\label{update}, \\
    \hspace*{-1em} w_{\boldsymbol{s},\boldsymbol{u},j} &= \frac{R_{\boldsymbol{s},\boldsymbol{u},j}}{\sum_{\boldsymbol{u}} R_{\boldsymbol{s},\boldsymbol{u},j}},
\end{align}
\end{subequations}
\end{small}
\par \noindent where  $T_{j}^{n}$ (resp. $R_{\boldsymbol{s},\boldsymbol{u},j}$) is the $j$-th component of the vector $\boldsymbol{T}^{n}$ (resp.~$\boldsymbol{R}_{\boldsymbol{s},\boldsymbol{u}}$). The detailed algorithmic procedure of the proposed method for decoupling and correcting scatter and beam-hardening artifacts in spectral CT is summarized in Algorithm \ref{Pseudo-code}.

\begin{algorithm}[ht] \label{Algorithm}
\caption{Pseudo-code of the proposed method}
\label{Pseudo-code}
\vspace{0.5em} 
\textbf{Input:} measured projection intensity $I^{t}$ \\
\textbf{Output:} final reconstructed RED image $\boldsymbol{x}^{*}$
\vspace{0.5em}

\begin{algorithmic}[1]
    \STATE \textbf{Initialize:} energy spectrum $q(E)$, detector response $D(E)$, and initial image $\boldsymbol{x}^{0}$ 
    
    \WHILE{not converged}
        \STATE \textbf{Step 1:} \textit{Estimate scatter components based on  $\boldsymbol{x}^{n}$}
        \SubState Calibrate scatter parameters $\boldsymbol{\phi}^{n}$ 
        \SubState Estimate multiple scattering $I^{s,n}_{\text{mul}}$ by \cref{multiple scattering}
        \SubState Estimate first-order scattering $I^{s,n}_{\text{fir}}$ by \cref{firOrder_scattering}
        
        \STATE \textbf{Step 2:} \textit{Estimate primary transmission}
        \SubState Estimate primary transmission $I^{p,n+1}_{\text{sc}}$ by \cref{corrPrimary}
        \SubState Estimate primary projection $P^{p,n+1}_{\text{sc}}$ by \cref{corrProj}
        
        \STATE \textbf{Step 3:} \textit{Update the image}
        \SubState Update the reconstructed image $\boldsymbol{x}^{n+1}$ by \cref{update}
        
        \STATE \textbf{Step 4:} \textit{Check for convergence}
       \SubState Check convergence criterion 
       \SubState $ n \leftarrow n + 1 $
    \ENDWHILE
    \STATE $\boldsymbol{x}^{*} \leftarrow \boldsymbol{x}^{n}$
    
\end{algorithmic}
\end{algorithm}

\section{Materials \& Experiments}
In this section, we present experiments evaluating the performance of our proposed model for the simultaneous correction of scatter and beam-hardening artifacts in spectral CT. The validation was conducted through extensive numerical simulations and physical experiments. Simulations were performed using the open-source MC-GPU software, while physical experiments were carried out on our in-house developed CT system. 
\subsection{Numerical simulations}
\subsubsection{Numerical phantoms}
We performed two sets of numerical simulations using phantoms with increasingly complex scatter distributions to evaluate the robustness of our proposed method. The first was a phantom designed to assess the algorithm's accuracy in material differentiation across a wide range of tissue densities. The second was an anatomically complex phantom to test the method's performance in a clinically realistic scenario with highly heterogeneous scatter distributions.
\begin{itemize}
    \item \textbf{Phantom 1: modified Gammex phantom.} 
    This numerical phantom consisted of a 20 cm diameter, 10 cm high water-filled cylinder (1.00 g/cm\textsuperscript{3}) containing several 3 cm diameter inserts to simulate a range of tissue densities. These inserts included materials representing lung tissue (0.385 g/cm\textsuperscript{3}), adipose (0.95 g/cm\textsuperscript{3}), dense trabecular bone (1.45 g/cm\textsuperscript{3}), and mineral bone (1.92 g/cm\textsuperscript{3}).
    \item \textbf{Phantom 2: Kyoto Kagaku Head phantom.} 
    This numerical phantom was digitized from a physical Kyoto Kagaku Head phantom ($24~\text{cm}\times 24~\text{cm} \times 23.6$~\text{cm}) to ensure clinical relevance. A scatter-free CT scan of the physical phantom provided a ground truth volume, which was voxelized to create a model with accurate geometry and attenuation for direct use in MC-GPU simulations.
\end{itemize}

\subsubsection{Simulation setups}
The detailed parameters for our MC simulations are provided in \cref{tab:MC_configurations}. Leveraging the low-frequency characteristics of scatter to mitigate computational demands, we downsampled the detector resolution from 2048 $\times$ 2048 to $256 \times 256$. This process correspondingly increased the size of each detector element from 0.02 cm $\times$ 0.02 cm to 0.16 cm $\times$
0.16 cm. The incident X-ray spectrum was generated using the SpekPy v2.0 software toolkit \cite{poludniowski2021spekpy}.

\begin{table}[htbp]
    \centering
    \caption{SCANNING CONFIGURATIONS OF MC SIMULATIONS}
    \label{tab:MC_configurations}
    \footnotesize
    \renewcommand{\arraystretch}{1.3} 
    \begin{tabular*}{\columnwidth}{@{} l @{\extracolsep{\fill}} c @{}}
        \hline\hline
        Configurations         & Value  \\ 
        \hline
        Tube voltage (kV)             & 120                 \\
        Filter                   & 1 mm Al / 6 mm Al       \\
        Detector array        & $256 \times 256$              \\
        Detector element size (cm\textsuperscript{2}) & $0.16 \times 0.16$       \\
        SOD\textsuperscript{a} (cm) & 80                \\
        SDD\textsuperscript{b} (cm) & 110                \\
        Number of projections           & 720                     \\
        Histories per projection    & $1\times 10^{11} $     \\
        \hline
        \multicolumn{2}{l}{\textsuperscript{a}Distance from X-ray source to the center of rotation.} \\
        \multicolumn{2}{l}{\textsuperscript{b}Distance from X-ray source to the detector.} \\
        \hline\hline
    \end{tabular*}
\end{table}

\subsubsection{Evaluation metrics}
To comprehensively evaluate the performance of different methods, we conducted a quantitative assessment from two perspectives: the accuracy of the scatter estimation and the quality of the final reconstructed images. For the first perspective, the scatter-to-primary-weighted mean absolute percentage error (SPMAPE) was employed to quantify the accuracy of the estimated scatter distributions \cite{erath2021deep}. For the second, the root mean square error (RMSE), the mean absolute error (MAE), and the mean relative error (MRE) were used to assess the quality of the reconstructed images with scatter correction.

\subsubsection{Benchmark}
To comprehensively validate the necessity and accuracy of our proposed method for the simultaneous correction of scatter and beam-hardening artifacts, its performance was benchmarked against several key references. These include an artifact-free ground truth image, the uncorrected reconstruction, and results from state-of-the-art methods such as IBHC \cite{krumm2008reducing}, PMSC \cite{gong2017physics}, FASKS \cite{sun2010improved}, and P-PolySKS (Polyquant-PolySKS) \cite{mason2018quantitative}.

\subsection{Experiments}
\subsubsection{Experimental setups}
To validate the effectiveness of our proposed method on real data, we conducted physical experiments on an in-house CBCT system. The system was equipped with a Powersite PSM-DT X-ray source and a Varian PerkinElmer XRD 1620 XN CS flat-panel detector. The first set of experiments utilized a Yin-Yang phantom (15~cm diameter and 6~cm high). This phantom was composed of two materials: a water-equivalent material ($1.0~\text{g/cm}^3$) and a bone-equivalent material ($1.92~\text{g/cm}^3$). 
The second set of experiments employed an anthropomorphic knee phantom composed of water-equivalent material and bony structures. 
\begin{figure}[pos=h] 
\centering
    \includegraphics[width=\linewidth]{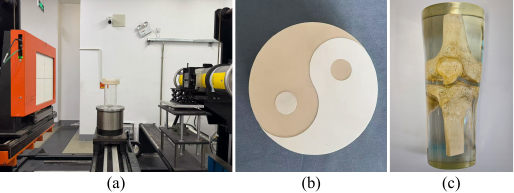}
    \caption{Experimental setup and phantoms. (a) In-house CBCT system; (b) Yin-Yang phantom; (c) Anthropomorphic knee phantom.}
    \label{Equipment}
\end{figure}
\cref{Equipment} shows a photograph of the experimental setup and the phantom used in the physical experiments. \Cref{tab_physical_parameters} summarizes the detailed scanning parameters used in this experiment. 
To establish a high-quality ground truth, an additional set of fan-beam projections was acquired for each phantom by tightly collimating the X-ray source. This geometry physically minimizes scatter, yielding nearly scatter-free data. The image reconstructed from these projections, though largely free of scatter artifacts, contained residual beam hardening. This artifact was corrected using our proposed BHC module to produce the final, artifact-free ground truth for all evaluations.

\begin{table}[htbp]
    \centering
    \caption{SCANNING PARAMETERS FOR THE EXPERIMENTS}
    \label{tab_physical_parameters} 
    \footnotesize
    \renewcommand{\arraystretch}{1.3}
    
    \begin{tabular*}{\columnwidth}{@{} l @{\extracolsep{\fill}} c c @{}} 
        \hline\hline
        
        Parameter                & Yin-Yang phantom & Knee phantom \\ 
        \hline
        Tube voltage (kV)       & 120              & 120              \\
        Tube current (mA)        & 4                & 1.6                \\
        Exposure time (s)        & 0.13             & 0.2             \\
        Filter                   & 1 mm Al          & 1 mm Al          \\
        Detector array (pixels)  & $2048 \times 2048$ & $2048 \times 2048$ \\
        Detector pixel size (cm\textsuperscript{2}) & $0.02 \times 0.02$ & $0.02 \times 0.02$ \\ 
        SOD\textsuperscript{a} (cm) & 88.5             & 115             \\
        SDD\textsuperscript{b} (cm) & 140              & 140              \\
        Number of projections    & 720              & 720              \\
       
        \hline
        
        \multicolumn{3}{l}{\textsuperscript{a}Distance from X-ray source to the center of rotation.} \\
        \multicolumn{3}{l}{\textsuperscript{b}Distance from X-ray source to the detector.} \\
        \hline\hline
    \end{tabular*}
\end{table}

\section{Results}
\subsection{Numerical simulations}
\begin{figure}[pos=h]
    \centering
    \includegraphics[width=\linewidth]{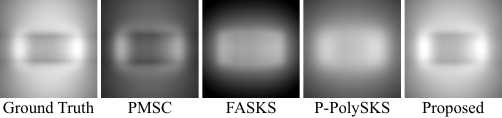}
    \caption{Comparison of the scatter distributions estimated by different methods for the modified Gammex phantom (random projection angle). All images are displayed in a window of [3000, 17000] photons.}
    \label{estimatedSca_Gammex}
\end{figure}

\begin{figure*}[htbp]
    \centering
    \includegraphics[width=\linewidth]{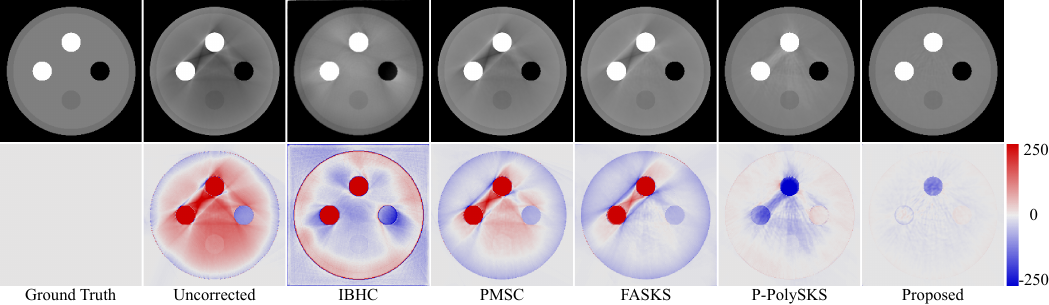}
    \caption{Quantitative comparison of reconstruction results for the central slice of the modified Gammex phantom. The top row shows the reconstructed images from different correction methods, displayed in a window of [-500, 500] HU. The bottom row displays the corresponding residual maps with respect to the ground truth, shown in a window of [-250, 250] HU. }
    \label{Recon_Gammex}
\end{figure*}

Given the structural uniformity of this Gammex phantom, the scatter distribution exhibits minimal variation across different projection angles. \cref{estimatedSca_Gammex} presents the estimated scatter profiles from various methods at a random projection angle. Following this, the corresponding corrected CBCT reconstruction results are shown in \cref{Recon_Gammex}. While several existing correction methods offer partial improvements, they still exhibit discernible residual errors. For instance, IBHC, PMSC, and FASKS systematically underestimate the bone inserts while overestimating the water regions, whereas P-PolySKS introduces overcorrection in the bone areas. In contrast, our method achieves superior performance, yielding a nearly artifact-free reconstruction with excellent agreement to the ground truth, except for a minor deviation in the densest bone. 
\begin{table}[htbp]
    \centering
    \caption{QUANTITATIVE EVALUATION OF DIFFERENT CORRECTION METHODS FOR THE GAMMEX PHANTOM}
    \label{tab_quantitative_gammex}
    \footnotesize %
    \renewcommand{\arraystretch}{1.3} %
    \begin{tabular*}{\columnwidth}{@{\extracolsep{\fill}} l c c c c}
        \hline\hline 
        \textbf{Method} & \textbf{SPMAPE}\textsubscript{$\downarrow$} & \textbf{RMSE}\textsubscript{$\downarrow$} & \textbf{MAE}\textsubscript{$\downarrow$} & \textbf{MRE (\%)}\textsubscript{$\downarrow$} \\
        \hline 
        Uncorrected     & 0.3565         & 0.0461                 & 0.0206                & 7.7349 \\
        IBHC            & ---            & 0.0476                 & 0.0194                & 7.0041 \\
        PMSC            & 0.1158         & 0.0299                 & 0.0130                & 4.9111 \\
        FASKS           & 0.0396         & 0.0207                 & 0.0105                & 4.2622 \\
        P-PolySKS         & 0.0273         & 0.0102                 & 0.0041                & 1.5603 \\
        \textbf{Proposed} & \textbf{0.0180} & \textbf{0.0034}       & \textbf{0.0015}       & \textbf{0.6293} \\
        \hline\hline 
    \end{tabular*}
\end{table}

\begin{table}[htbp]
    \centering
    \caption{QUANTITATIVE EVALUATION OF DIFFERENT CORRECTION METHODS FOR THE HEAD PHANTOM}
    \label{tab_quantitative_headPhantom}
    \footnotesize 
    \renewcommand{\arraystretch}{1.3} 
    \begin{tabular*}{\columnwidth}{@{\extracolsep{\fill}} l c c c c}
        \hline\hline 
        \textbf{Method} & \textbf{SPMAPE}\textsubscript{$\downarrow$} & \textbf{RMSE}\textsubscript{$\downarrow$} & \textbf{MAE}\textsubscript{$\downarrow$} & \textbf{MRE (\%)}\textsubscript{$\downarrow$} \\
        \hline 
        Uncorrected     & 0.5177         & 0.0608                 & 0.0361                & 11.9669 \\
        IBHC            & ---            & 0.0606                 & 0.0372                & 11.7358 \\
        PMSC            & 0.1792         & 0.0396                 & 0.0218                & 7.1286  \\
        FASKS           & 0.0633         & 0.0257                 & 0.0155                & 5.4894  \\
        P-PolySKS         & 0.0368         & 0.0162                 & 0.0087                & 3.4735  \\
        \textbf{Proposed} & \textbf{0.0083} & \textbf{0.0068}       & \textbf{0.0037}       & \textbf{1.2744} \\
        \hline\hline 
    \end{tabular*}
\end{table}
\begin{figure}[pos=htbp]
\centering
    \includegraphics[width=\linewidth]{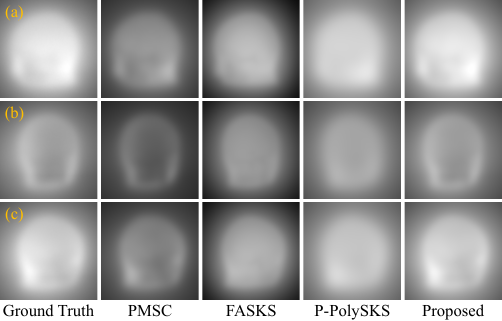}
    \caption{Visual comparison of the scatter distributions estimated by different methods for the head phantom. The rows correspond to three distinct projection views: (a) 0\textdegree, (b) 120\textdegree, and (c) 150\textdegree. The display window for all images is set to [1500, 30000] photons.}
    \label{estimatedSca_Head}
\end{figure}
\begin{figure*}[htbp]
    \centering
    \includegraphics[width=\linewidth]{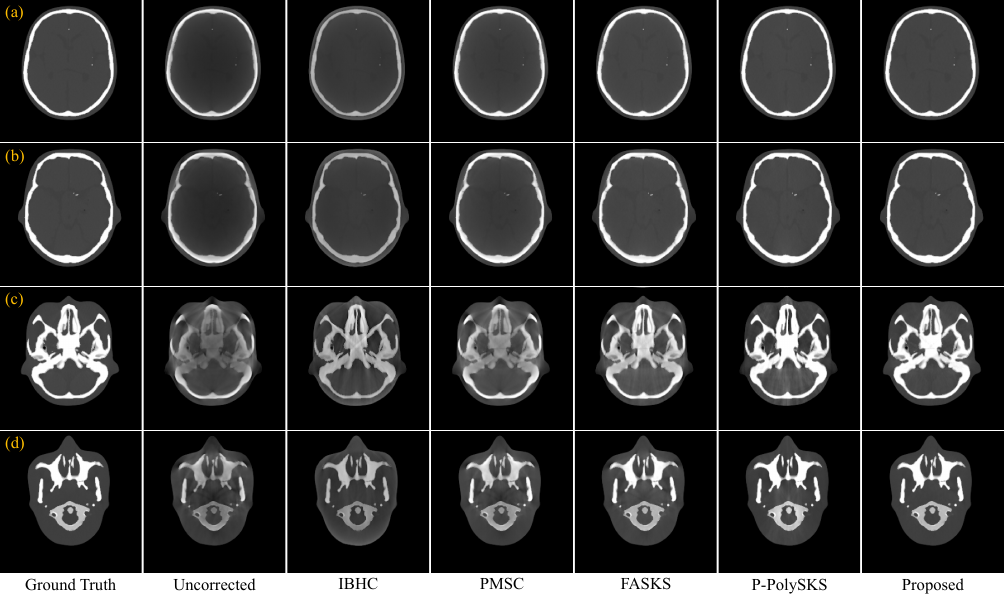}
    \caption{Reconstructed images of the head phantom at different axial levels: (a-b) posterior cranial regions and (c-d) mid-facial regions. All images are displayed in a window of [-500, 1500] HU. }
    \label{Recon_Head}
\end{figure*}

Quantitative results for the Gammex phantom are summarized in \Cref{tab_quantitative_gammex}. In scatter estimation, our method achieves the lowest SPMAPE, outperforming all comparison methods and corroborating the visual advantage observed in \cref{estimatedSca_Gammex}. For overall image quality, our approach outperforms all others across all metrics, improving accuracy by approximately an order of magnitude in RMSE, MAE, and MRE compared with the uncorrected case.

Unlike the structurally uniform Gammex phantom, the head phantom (the corresponding reconstructions are compared in \cref{Recon_Head}) exhibits significant view-dependent variations in its scatter distribution, as shown in \cref{estimatedSca_Head}.  
In the posterior cranial slices, severe scatter-induced cupping artifacts in the uncorrected image obscure internal brain tissue details. 
All correction methods substantially mitigate these artifacts, with FASKS, P-PolySKS, and the proposed method yielding results visually nearly indistinguishable from the ground truth. 
Performance disparities become markedly more pronounced in the complex midface regions. While the PMSC method leaves residual visual artifacts in bone and soft tissue, other comparison methods (IBHC, FASKS, P-PolySKS) introduce new streak artifacts at high-density bone edges (high SPR regions) and exhibit incomplete correction in soft tissues. 
In contrast, the proposed method demonstrates exceptional artifact suppression capability in these challenging regions.
These qualitative findings are supported by the quantitative results summarized in \cref{tab_quantitative_headPhantom}, which exhibit a trend highly consistent with the Gammex phantom results (\cref{tab_quantitative_gammex}).
\begin{figure*}[ht]
    \centering
    \includegraphics[width=\linewidth]{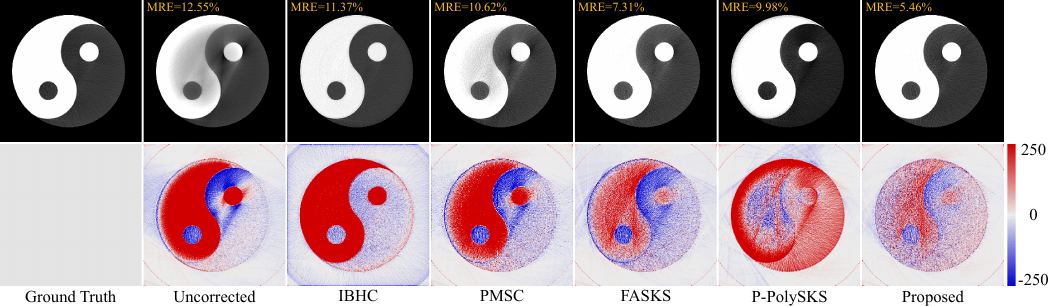}
    \caption{Qualitative and quantitative comparison of reconstruction results for a slice of the physical Yin-Yang phantom. The top row shows the reconstructed images from various methods, annotated with their respective MRE values. All reconstructed images are shown in a window of [-200, 1200] HU. The bottom row presents the corresponding difference maps relative to the fan-beam ground truth, shown in a window of [-250, 250] HU. }
    \label{Recon_Real_YinYang15cm}
\end{figure*}
\begin{figure*}[ht]
    \centering
    \includegraphics[width=\linewidth]{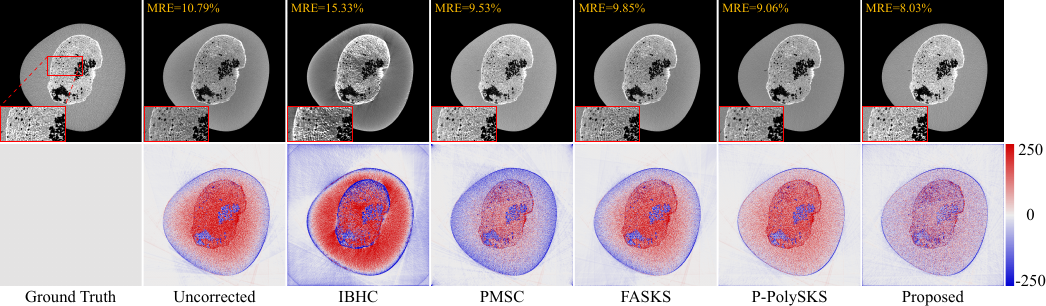}
    \caption{Qualitative and quantitative comparison of reconstruction results for a slice of the physical knee phantom. The top row shows the reconstructed images from various methods, annotated with their respective MRE values. All reconstructed images are displayed in a window of [-500, 500] HU. The bottom row presents the corresponding difference maps relative to the fan-beam ground truth, shown in a window of [-250, 250] HU.}
    \label{Recon_Real_Knee30cm}
\end{figure*}
\subsection{Experiments}
\cref{Recon_Real_YinYang15cm} presents the reconstruction results for the Yin-Yang phantom. All methods reduce artifacts compared to the uncorrected case, but with key differences. PMSC leaves noticeable residual artifacts, while P-PolySKS shows errors primarily at the phantom boundary. Although IBHC produces a visually uniform image, its difference map reveals severe underestimation within the bone region. FASKS performs well overall but exhibits inaccuracies in the central region. In contrast, the proposed method achieves optimal correction, yielding a reconstruction nearly indistinguishable from the ground truth, attaining the lowest MRE, and producing a difference map that confirms minimal residual error at the water–bone interface.
\begin{figure*}[pos=h]
    \centering
    \includegraphics[width=\linewidth]{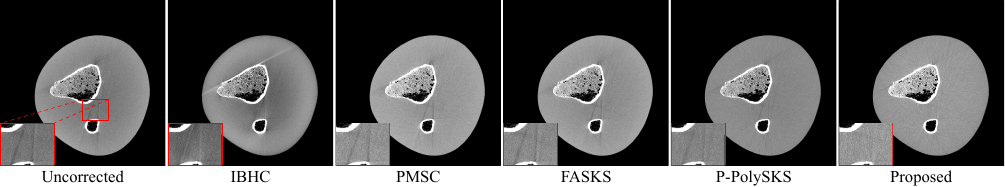}
    \caption{Visual comparison of reconstruction results for a representative axial slice of the physical knee phantom. All images are displayed in a window of [-500, 500] HU.}
    \label{Recon_Real_Knee30cm_NoSlit}
\end{figure*}

\cref{Recon_Real_Knee30cm} presents the reconstruction results for the central slice of the knee phantom, highlighting trabecular bone details. The uncorrected image exhibits severe cupping in soft tissue and degraded contrast of trabecular structures. While both FASKS and P-PolySKS suffer from scatter underestimation, resulting in shading artifacts, the IBHC method introduces additional  streak artifacts.
The results from the PMSC method and our proposed method are the most visually comparable, as both effectively restore the bone details. However, in the soft-tissue areas and within the non-uniform void regions of the bone, the correction by PMSC remains inaccurate. In contrast, our method provides a more thorough and accurate scatter correction in these challenging regions. 
This qualitative observation is further substantiated by the quantitative MRE metrics, which confirm the superior performance of our approach. 

To further demonstrate the robustness of our method, \cref{Recon_Real_Knee30cm_NoSlit} presents the reconstruction results at an off-center axial slice (without an ideal scatter-free ground truth). The uncorrected image and its magnified region of interest (red box) exhibit a combination of both shading and streak artifacts.  While the FASKS and PMSC methods attenuate the shading artifacts to some extent, they have a negligible effect on the streak artifacts. The P-PolySKS method performs slightly better, showing some suppression of both artifact types, yet faint residual artifacts are still perceptible. In contrast, the result from our proposed method demonstrates clear superiority. From the global structure to the fine local details, virtually no residual artifacts are visually perceptible.

\section{Discussion}
\subsection{Necessity of joint decoupling}
Both X-ray scatter and beam-hardening artifacts manifest with visually similar characteristics, such as cupping and streaking, in reconstructed CBCT images, despite originating from distinct physical phenomena. A common pitfall arises when artifact correction strategies solely address beam hardening while neglecting the contribution of scattered radiation. 
For instance, as illustrated in \cref{Recon_Gammex,Recon_Head}, IBHC approach often introduces additional radial artifacts in the vicinity of high-attenuation structures. Furthermore, as shown in \cref{Recon_Real_YinYang15cm}, even if the reconstruction appears visually artifact-free, failing to account for scatter can depress the quantitative values within the reconstructed image. 
This phenomenon might lead to an ostensibly ideal reconstruction that, in reality, provides misleading quantitative information. Consequently, establishing an accurate forward model for CBCT imaging that can effectively decouple and correct for both scatter and beam-hardening artifacts is important. 

\subsection{Validation of decoupling accuracy}
\begin{figure}
\centering
    \includegraphics[width=\linewidth]{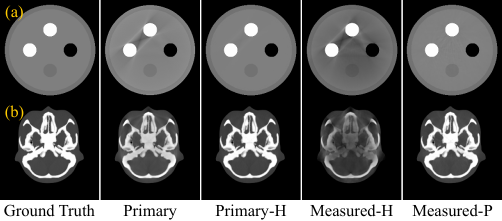}
    \caption{Ablation study results demonstrating the decoupling accuracy of the proposed method. (a) Gammex phantom slice ([-500, 500] HU). (b) Anthropomorphic head phantom slice ([-500, 1500] HU). Five scenarios are compared: Ground Truth (ideal reference); Primary (scatter-free projections, highlighting intrinsic beam hardening); Primary-H (scatter-free projections with the proposed BHC, verifying BHC accuracy); Measured-H (measured data with BHC only, showing that residual scatter compromises BHC effectiveness); and Proposed (full decoupling and correction on measured data).}
    \label{Discussion_Ablation}
\end{figure}

Given the inherently similar visual characteristics of scatter and beam-hardening artifacts, a rigorous ablation study is essential to validate the effective decoupling by our method. 
The results of this study are presented in \cref{Discussion_Ablation} and analyzed as follows in three sequential steps. 
First, to validate the performance of the BHC module, we compare the Ground Truth with the Primary reconstruction.
Since the Primary CBCT image is reconstructed from scatter-free primary projections, the artifacts present within it can be purely attributed to the beam-hardening effect. Subsequently, the Primary-H CBCT image is reconstructed from primary projections. The reconstruction result is nearly indistinguishable from the Ground Truth, which validates the accuracy of our BHC module. 
Second, to demonstrate that scatter correction and BHC are not interchangeable, we compare the Primary-H and Measured-H reconstructions.
The Measured-H CBCT image is reconstructed from scatter-contaminated measured projections with the BHC module. 
In contrast to the Primary-H result, the Measured-H result is corrupted by severe cupping and shading artifacts, bearing a strong resemblance to the uncorrected reconstructions shown in \cref{Recon_Gammex,Recon_Head}.
This comparison further demonstrates that the presence of uncorrected scatter severely degrades the performance of the BHC module.
Finally, we compare the Measured-H result with that of our proposed method. The proposed method 
successfully eliminates the artifacts present in Measured-H and recovers an image quality that is highly consistent with both the Ground Truth and the Primary-H result. This final comparison confirms that our proposed method achieves an accurate decoupling and simultaneous correction of both scatter and beam-hardening artifacts. 
\subsection{Comparative convergence analysis}
\begin{figure}
    \centering
    \includegraphics[width=\linewidth]{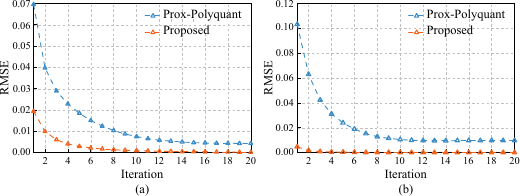}
    \caption{Comparison of RMSE convergence curves between the two BHC methods, evaluated on (a) the slice of the modified Gammex phantom, and (b) the midface slice of the anthropomorphic head phantom.}
    \label{Convergence_Speed}
\end{figure}

To assess the efficiency of the proposed voxel-adaptive BHC module, we conducted a quantitative comparison against the Prox-Polyquant baseline, focusing on the convergence speed and reconstruction accuracy.
We selected a slice of the modified Gammex phantom (shown in \cref{Recon_Gammex}) and a midface slice of the anthropomorphic head phantom (corresponding to \cref{Recon_Head}(c)) as test objects. \cref{Convergence_Speed} illustrates the RMSE convergence curves for the two methods, where subplots (a) and (b) correspond to results of the Gammex phantom and the head phantom, respectively. As shown in the figure, the proposed method demonstrates significantly faster convergence in the early stages for both phantoms. 
Specifically, at the 3rd iteration, the RMSE of our method drops to 0.00600 and 0.00090 for the Gammex and head phantoms, respectively, whereas the Prox-Polyquant baseline remains at higher levels of 0.02890 and 0.04240. Furthermore, after 20 iterations, the proposed BHC module stabilizes at a lower RMSE level compared to the baseline for both cases. Quantitative results indicate that for the Gammex and head phantoms, the final RMSE values of our method are 0.00014 and 0.00001, respectively, which are notably lower than the baseline values of 0.00420 and 0.00970. These results further confirm that the adaptive BHC module not only significantly accelerates early convergence but also effectively improves the final accuracy and fidelity of the reconstructed image. 
\subsection{Current limitations and strategies}
This study has achieved advancements in decoupling and correcting both scatter and beam hardening artifacts. Nevertheless, we observed that when determining the convolution kernel parameters for multiple-scattering using the least squares method, the choice of initial parameters impacts the correction efficacy. To circumvent the tediousness of manual parameter tuning, we propose to make use of the smooth, low-frequency characteristics of multiple-scattering by employing a shallow convolution network for learning its distribution. 
While a DL model could learn the entire scattering distribution, modeling the physically complex, high-frequency first-order scattering via a purely data-driven approach would demand a network with excessive parameters and computational resources. 
Conversely, the low-frequency nature of multiple scattering makes it amenable to effective learning by a shallower network.

\section{Conclusion}
X-ray scatter and beam hardening are two primary factors that critically degrade CBCT image quality. These distinct physical phenomena introduce coupled and visually similar artifacts, presenting a significant challenge for reconstructing CBCT images with high accuracy. To address this, we propose a more accurate CBCT forward process, built upon the Polyquant attenuation framework. Our approach introduces key innovations for both artifact sources. We first present a hybrid scatter correction strategy that uses a physics-based process to accurately compute high-frequency first-order scattering, while employing an object-adaptive convolution module for rapid estimation of multiple scattering. Concurrently, we develop a novel BHC module, which adaptively determines the optimal update for each pixel during each iteration. This mechanism not only eliminates the need for manual parameter tuning but also accelerates convergence and enhances reconstruction accuracy.



\section*{Acknowledgements}
We thank Dr. Jonathan H. Mason for generously making the PolySKS code open-source, which was essential for the scatter correction performed in this work. We are also grateful to Dr. Jigang Duan and Dr. Heran Wang for their assistance during the experimental data acquisition. 
\bibliographystyle{IEEEtran}
\bibliography{reference_CMPB}

\end{document}